\documentclass[12pt,preprint]{aastex}
\pdfoutput=1
\usepackage{emulateapj5}
\begin{document}
 
\newcommand{\kms}{\mbox{km~s$^{-1}$}}
\newcommand{\s}{\mbox{$''$}}
\newcommand{\mloss}{\mbox{$\dot{M}$}}
\newcommand{\my}{\mbox{$M_{\odot}$~yr$^{-1}$}}
\newcommand{\ls}{\mbox{$L_{\odot}$}}
\newcommand{\ms}{\mbox{$M_{\odot}$}}
\newcommand\mdot{$\dot{M}  $}
\title{High-Velocity Interstellar Bullets in IRAS05506+2414: A Very
Young Protostar}
 
\author{Raghvendra Sahai\altaffilmark{1}, Mark Claussen\altaffilmark{2}, Carmen S{\'a}nchez
Contreras\altaffilmark{3}, Mark Morris\altaffilmark{4}, and Geetanjali
Sarkar\altaffilmark{5}
}

\altaffiltext{1}{Jet Propulsion Laboratory, MS\,183-900, Caltech,
    Pasadena, CA 91109}
 
\altaffiltext{2}{National Radio Astronomy Observatory, 1003 Lopezville Road,
Socorro, NM 87801}
 
\altaffiltext{3}{Dpto. de Astrof\'{\i}sica Molecular e
Infraroja, Instituto de
Estructura de la Materia-CSIC, Serrano 121, 28006 Madrid, Spain}
 
\altaffiltext{4}{Division of Astronomy \& Astrophysics, UCLA, Los Angeles,
CA 90095-1547}

\altaffiltext{5}{Department of Physics, Indian Institute of Technology,
Kanpur-208016, U.P., India}
 
\email{raghvendra.sahai@jpl.nasa.gov}
 
\begin{abstract}
We have made a serendipitous discovery of an enigmatic outflow source, IRAS\,05506+2414
(hereafter IRAS\,05506), as part of a multi-wavelength survey of pre-planetary nebulae
(PPNs). The HST optical and near-infrared images show a bright compact central source
with a jet-like extension, and a fan-like spray of high-velocity (with radial velocities
upto 350\,\kms) elongated knots which appear to emanate from it. These structures are possibly
analogous to the near-IR ``bullets" seen in the Orion nebula. Interferometric
observations at 2.6\,mm show the presence of a high-velocity CO outflow and a continuum
source also with a faint extension, both of which are aligned with the optical jet
structure. IRAS\,05506 is most likely not a PPN. We find extended NH$_3$ (1,1) emission
towards IRAS\,05506; these data together with the combined presence of far-IR emission, H$_2$O
and OH masers, and CO and CS J=2--1 emission, strongly argue for a dense, dusty star-forming
core
associated with IRAS\,05506. IRAS\,05506 is probably an intermediate-mass or massive
protostar, and the very short time-scale (200\,yr) of its outflows indicates that it is
very young. If IRAS\,05506 is a massive star, then the lack of radio continuum and the
late G -- early K spectral type we find from our optical spectra implies that in this
object we are witnessing the earliest stages of its life, while its temperature is still
too low to provide sufficient UV flux for ionisation.
\end{abstract}

\keywords{ISM: clouds, ISM: individual objects: IRAS05506+2414, 
stars: formation, 
stars: pre-main sequence, stars: mass loss, radio lines: ISM }

\section{Introduction} We report the serendipitous discovery of an energetic outflow
source
from the object IRAS\,05506+2414 (hereafter IRAS\,05506). IRAS\,05506 was observed as
part of our multi-wavelength survey of pre-planetary nebulae (PPNs) (Sahai et al. 2007),
which are transition objects between the AGB and planetary nebula evolutionary phases in
the lives of low and intermediate mass (1-8\ms) stars. These objects were selected from OH
maser catalogs using the IRAS color criterion, $F_{25}/\,F_{12}\,>\,1.4$. Subsequent and
continuing observation of
the resulting catalog of PPN candidates has shown that the catalog contains PPNs almost
exclusively, but the object we discuss here is a notable exception.

In this paper, we report our HST imaging of IRAS\,05506 in the optical and near-infrared,
which has revealed a morphology
unlike that of any known PPN. We also present additional observations of optical, 
millimeter, and centimeter-wave lines as well as the radio continuum towards IRAS\,05506,
and discuss the possible nature of this source. The plan of our paper is as follows: in
\S\,2 we describe our observations, in \S\,3 we present the results of these observations,
in \S\,4 we derive the physical properties of IRAS\,05506 which show that it is most
likely not a PPN but a very young protostar, and in \S\,5 we summarize
our main conclusions.

\section{Observations}
\subsection{Optical and Near-Infrared Imaging with HST }
IRAS\,05506 (also AFGL\,5171) was imaged on UT date 2002 October 20 (GO program 9463), by
the High Resolution
Camera (HRC) of the Advanced Camera for Surveys (ACS), which has a plate scale of
0\farcs025/pixel, using the F606W filter ($\lambda=0.59\micron$,
$\Delta\lambda=0.157\micron$; exposures were 2$\times$400\,s) and the F814W filter
($\lambda=0.81\micron$, $\Delta\lambda=0.166\micron$; exposures were 2$\times$200\,s). All
images were obtained with a 2-point dither, and the standard STScI/HST pipeline
calibration
has been applied to all data. We also obtained near-infrared images on UT date 2004 March
23 (GO program 9801) with Camera 2 of the Near Infrared
Camera and Multi-Object Spectrometer (NICMOS), which has a plate scale of
0\farcs076/pixel (GO program
9801), using the F110W ($\lambda=1.1\micron$, $\Delta\lambda=0.38\micron$; total exposure
time was 544\,s), F160W ($\lambda=1.6\micron$, $\Delta\lambda=0.28\micron$; total exposure
time was 56\,s), and F205W ($\lambda=2.1\micron$, $\Delta\lambda=0.43\micron$; total
exposure time was 18\,s) filters. For ease of comparison, the
NICMOS images have been rotated to the same orientation as the HRC/ACS images. Numerous
field stars present in the images have been used to achieve satisfactory
registration between the optical and near-infrared images.

\subsection{Optical Spectroscopy}
We obtained long-slit spectra of IRAS\,05506 on 2004 Nov 8 with the Echellete Spectrograph
and Imager (ESI; Sheinis et al., 2002) on the 10\,m W.M. Keck\,II telescope at Mauna Kea
(Hawaii, USA). The detector was a MIT-LL CCD with
2048$\times$4096 squared pixels of 15$\mu$m. Total wavelength coverage
is $\sim$3900-10900\AA. The reciprocal dispersion and the pixel
angular scale range from 0.16 to 0.30 \AA/pixel and from 0\farcs120 to
0\farcs168, respectively, for the ten echellette orders (15 to 6) of
ESI, which was used in its echelle mode. The velocity dispersion has a
nearly constant value of 11.5\,\kms\,pixel$^{-1}$ in all orders.  We
used a 0\farcs5$\times$20\arcsec\ slit with its long side aligned along a vector which
passes through two of the most 
prominent emission knots seen towards this source, with an exposure time of 600\,s.  
Further technical details for these observations are
described in S{\'a}nchez Contreras et al. (2008, in prep) and 
are similar to those described by Sahai et al. (2005).

\subsection{Millimeter-Wave Interferometry}
The OVRO millimeter-wave array\footnote{\tt http://www.ovro.caltech.edu/mm}, in its low
(L) configuration (for which baselines
between antennas range from 15\,m to 115\,m), was used to obtain snapshot
observations of the CO J=1--0 line and the 2.6\,mm continuum towards IRAS\,05506 on 2003
Mar 17 and 24. 
The units of the digital spectral line correlator were arranged to
provide a total bandwidth of 90\,MHz ($\sim$\,234\,\kms) with a
channel spacing of 1\,MHz (corresponding to $\sim$2.6\,\kms).  The
3\,mm continuum emission was observed simultaneously using the
dual-channel analog continuum correlator, which provided a total
bandwidth of 4\,GHz (after combining both IF bands).  Total
integration time on source and calibrators was $\sim$\,11\,hr.

The calibration of the data was performed using the MMA software
package\footnote{MMA is written and maintained by the Caltech's
Millimeter Interferometry Group.}.  Data were gain calibrated in
baseline-based mode using the quasar J0530$+$135, which was observed
at regular time intervals of $\sim$\,20\,minutes before and after our
target. The quasars 3C\,273 and 3C\,84 were used
as passband and flux calibrators. 

Reconstruction of the maps from the visibilities was done using the
Multichannel Image Reconstruction, Image Analysis and Display (MIRIAD)
software. We Fourier transformed the measured visibilities with robust
weighting (which is an optimized compromise between natural and
uniform weighting) for CO and natural weighting for the continuum map for
S/N optimization. After that, the images obtained were cleaned and maps restored.
The (fitted Gaussian) clean beam for our CO (continuum) map has FWHM=$3.9{''}\times2.9{''}$
($4.6{''}\times3.2{''}$) and is oriented at PA=-23.6\arcdeg (-15.3\arcdeg). 
The 1$\sigma$ noise in our CO (continuum) maps, as measured in
channels with no signal, is 0.45 (1.1)\,mJy\,beam$^{-1}$. The
conversion factor from CO (continuum) surface brightness to
temperature units is 8.0 (6.3)\,K per Jy\,beam$^{-1}$.



\subsection{Centimeter Ammonia, Water and Methanol Maser Lines}
Observations of the H$_2$O, NH$_{3}$ and methanol lines in the K-band, i.e.
$(22-25)$\,GHz, were obtained with the NRAO\footnote{The NRAO (National Radio Astronomy
Observatory) is a facility of the National Science Foundation operated under cooperative
agreement by Associated Universities, Inc.} Green Bank Telescope (GBT) on February 28,
2006 and March 3, 2006. A 6$\times$6 raster scan (step-size=13$''$), centered on the
J2000 position RA = 05:53:43.6, Dec = +24:14:44.5 was performed in right ascension and
declination 4 times during these two days. The beam of the telescope was measured during
pointing scans to be 32$''$, so the raster was Nyquist sampled over a small region in the
sky of slightly more than $\sim$1 arcminute. A dual-beam receiver was used at
(22--25)\,GHz, and total power measurements were made, relative to an off-source position
of RA = 06:02:29.96, Dec = +26:14:44.2 (J2000). The GBT spectrometer backend was used to
record the total-power spectra in four different 50\,MHz frequency bands. These four
bands were set to Doppler track the 22235.08 H$_2$O line ($6_{16}\rightarrow5_{23}$), the
average of the NH$_3$ (1,1) and (2,2) metastable inversion lines at 23694.50 MHz and
23722.63 MHz, respectively, the NH$_3$ (3,3) metastable inversion line at 23870.1292 MHz,
and the three methanol lines at $\sim$24932.0 MHz
($3_{2,1}\rightarrow3_{1,2}\,E$,$4_{2,2}\rightarrow4_{1,3}\,E$,
$2_{2,0}\rightarrow2_{1,1}\,E$). The 50\,MHz bands provided velocity coverage of
$\sim$650\,\kms, centered at $V_{LSR}$=0\,\kms. Each 50 MHz frequency band had 8192
spectral channels, providing a spectral resolution of 6.1\,kHz ($\sim$0.08\,\kms). For
each raster position, a total integration time of $\sim$16 minutes was obtained,
providing an rms antenna temperature of $\sim$25 mK.

The VLA was used to observe the H$_2$O line on June 14, 2005, in the CnB
configuration (beam=$1.10{''}\times0.70{''}$ arcseconds at PA=55\arcdeg), with 
the correlator configured to provide 127 spectral channels ($\Delta\nu$=
48.8\,kHz) across a bandwidth of 6.25 MHz (84\,\kms). Three velocity ranges, centered at
$V_{LSR}$=$-70$, 0, and $+70$\,\kms~(thus covering a total velocity range from
$-$112 to $+$112\,\kms) were searched by time multiplexing 
the observations. Each velocity setting was observed for about 10 minutes.
The data were calibrated and imaged in the standard way using the
NRAO's Astronomical Imaging Processing System (AIPS). The rms noise in the image of
each spectral channel was $\sim$15 mJy beam$^{-1}$.

\subsection{Radio Continuum Observations}
We observed the radio continuum at 8.4\,GHz (3.6\,cm) in IRAS\,05506 on June 11, 2005,
using the VLA in the CnB configuration. Approximately 13 minutes were spent on source.
The data was calibrated in the usual manner, using AIPS. A continuum image was made using
the calibrated {\it u, v} data; the
synthesized beam, using natural weighting, was $5\farcs1\times1\farcs1$ at a position
angle of 56$\arcdeg$. The rms noise in the image was $\sim$50 $\mu$Jy beam$^{-1}$,
similar to the expected theoretical noise.

We also examined the region around the position of IRAS\,05506 in the NRAO VLA Sky Survey
(NVSS; Condon et al. 1998). The NVSS observed at a frequency of 1.4\,GHz (21\,cm), and
typically had noise levels in less-confused regions of about 400 $\mu$Jy beam$^{-1}$. This
was approximately the noise in the field containing the position of IRAS\,05506.

%


\section{Results}
\subsection{Optical and Near-Infrared Imaging}
\begin{figure}[htbp]
\vskip -0.5cm
\resizebox{1.0\textwidth}{!}{\includegraphics{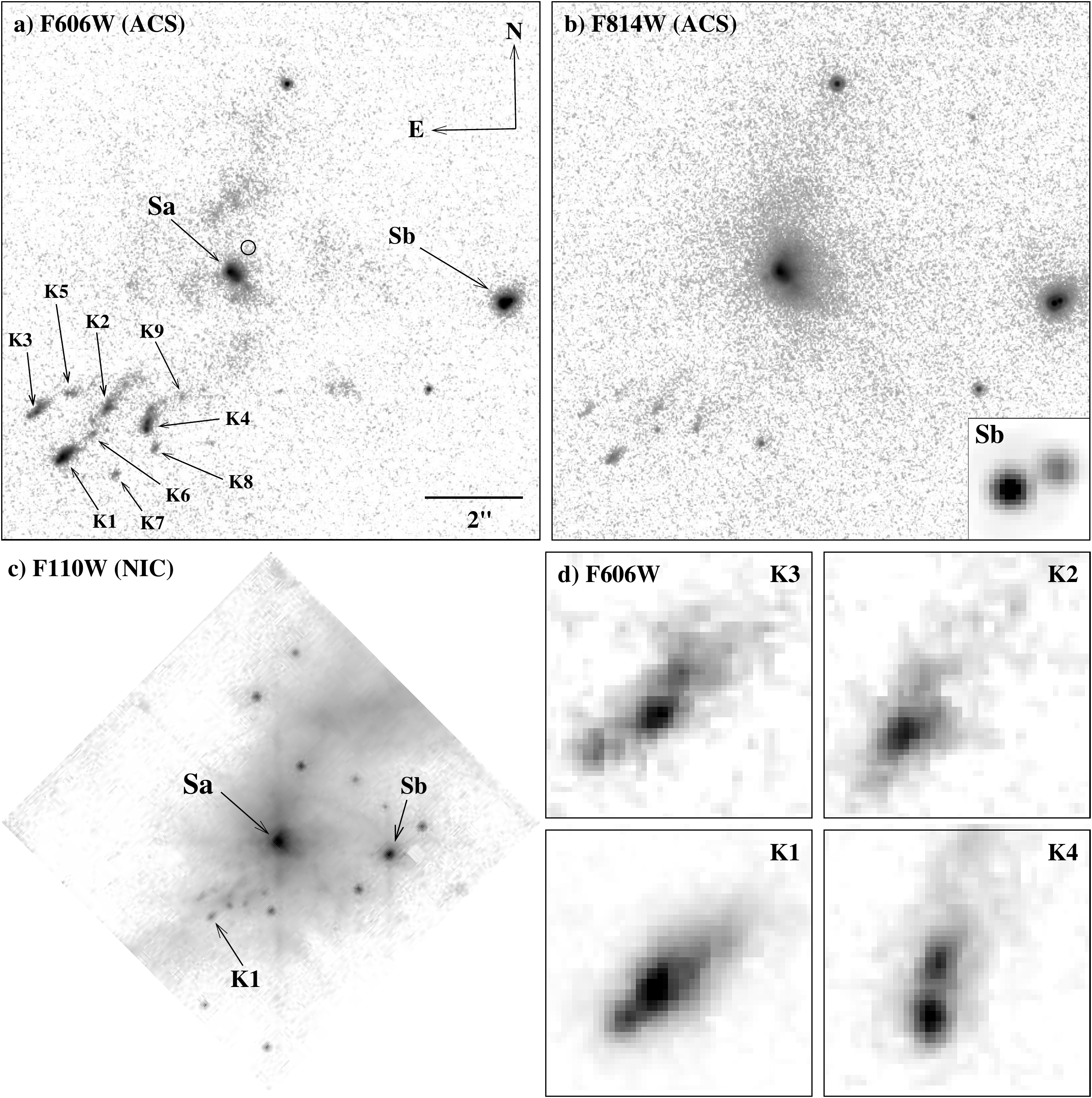}}
\vskip 0.5cm
\caption{HST/ACS images of IRAS\,05506. Panels a \& b show the optical images taken with
ACS ({\it log stretch}). The position (J2000 RA = 05:53:43.54, Dec =
24:14:45.2) of the $H_2$O maser emission feature detected with the VLA is shown as a small
circle in panel a; the size of the circle denotes the positional uncertainty. An expanded
view of source $S_b$ (inset, panel b) shows a close binary pair with a separation of
0\farcs12. Panel c shows the F110W taken with NICMOS ({\it log stretch}) -- a small square
patch to the right of {\it Sb} corrupted by an instrumental artifact has been blanked out. A
few of the features in the panel a are also labelled in c for visual registration. Panel {\it
d}
shows expanded views (0\farcs63$\times$0\farcs63) of the knots $K1-K4$ extracted from the F606W
image ({\it linear stretch}).
}
\label{hst606}
\end{figure}

\begin{figure}[htbp]
\vskip -0.5cm
\resizebox{0.85\textwidth}{!}{\includegraphics{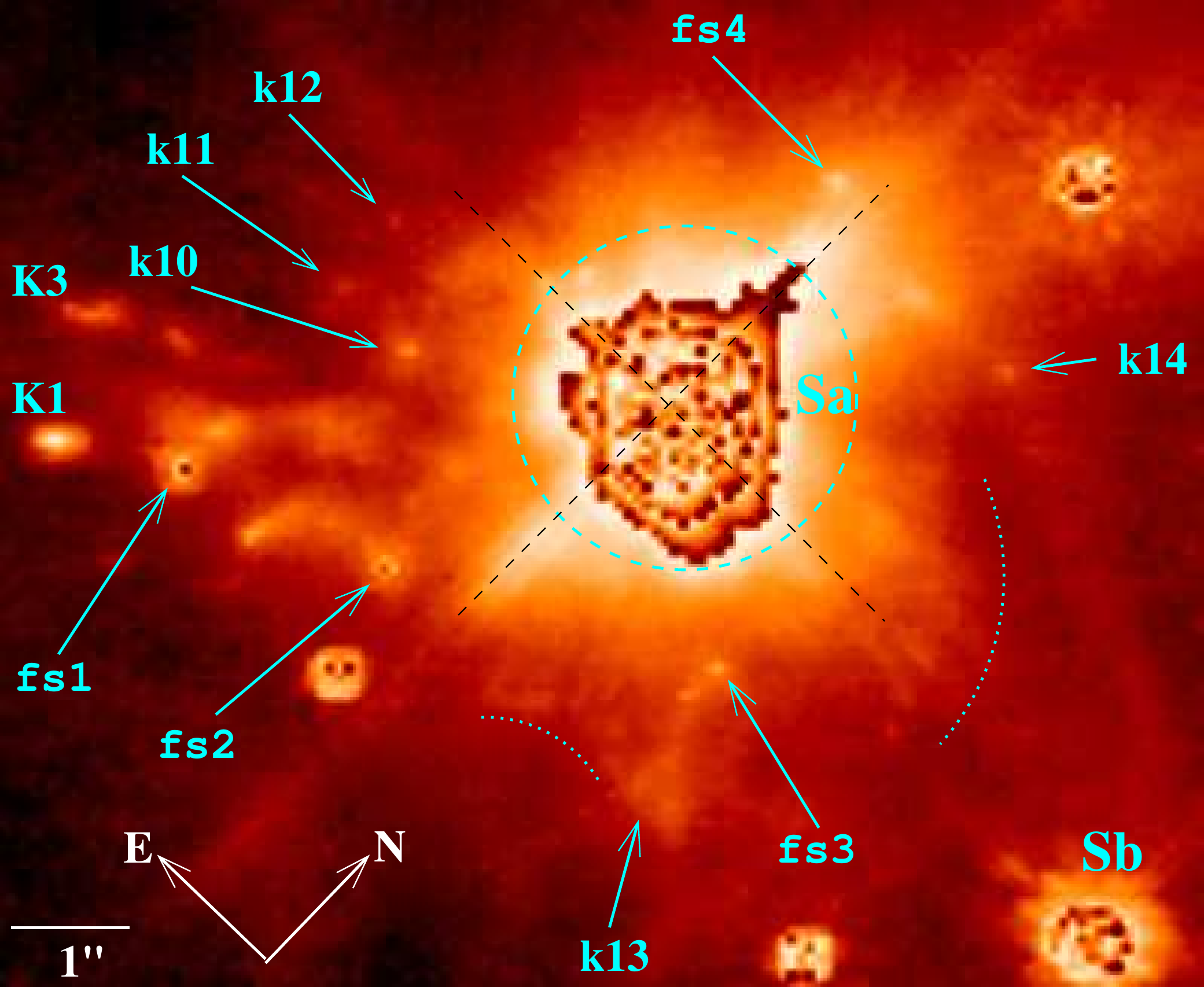}}
\vskip 0.5cm
\caption{F110W NICMOS image of IRAS\,05506 highlighting the presence of knotty structures
($k10-k14$) not seen in the F606W ACS image (some of the common features ($K1, K2, Sa$, and
$Sb$ are also labelled for visual registration). The greyscale shows $log(log(I))$, where $I$
is the intensity; above a threshold value, the color stretch cycles repeatedly
through the range of colors utilized, giving the mottled appearance to the brightest regions in
the image. Faint stars in this image are labelled $fs1-fs4$. The dashed circle covers a
central region around {\it Sa} where knotty structures cannot be reliably distinguished from
PSF structures. The dotted arcs delineate angular wedges where faint filamentary structures
can be seen; many of these appear to be radially directed away from {\it Sa}. The spiky linear
structures oriented at $\pm$45\arcdeg~to the horizontal (denoted by dashed black lines)
which appear to emanate from {\it Sa} are diffraction spikes produced by the telescope optics.
Note that the
orientation of this image is different from the one in
Fig.\,\ref{hst606}. 
}
\label{jband2}
\end{figure}

The HST F606W image (Fig.\ref{hst606}a) shows a compact source ({\it Sa}), and a fan-like spray
of compact nebulous features $K1-K9$ that are separated from {\it Sa}, but appear to emanate
radially from it within a range of position angles (121\arcdeg--155\arcdeg). Examination of
the most prominent of these knots ($K1-K4$) shows that each of them is a highly collimated
structure. Most of these knots are seen in the other filters used for the HST imaging as well.
A few more knots ($k10-k12$), not seen in the optical images, become visible in the F110W and
F160W images in the PA range 85\arcdeg--121\arcdeg (Fig.\,\ref{jband2})\footnote{in this
figure, we have retained
the original orientation of the NICMOS images (which is different from the optical images
taken with the ACS) in order to avoid any loss of fidelity which would result from image
rotation}, at locations closer
to {\it Sa}, presumably because these are more highly extincted than the $K1-K4$ knots.
Compact knots like $k10$ can be distinguished from faint stars in the NICMOS images (labelled
$fs1-fs4$ in Fig.\,\ref{jband2}) because, relative to their
surroundings, the knots are brightest in the F110W image (i.e., at the shortest IR wavelength),
whereas the stars are brightest
in the F205W images (i.e., at the longest IR wavelength). Significant filamentary structures
can be seen all around {\it Sa}. The most prominent of these lie in the western hemisphere
around {\it Sa}, and several of these appear to be radially directed away from the latter
(dotted arcs in Fig\,\ref{jband2} delineate angular wedges where these structures are most
clearly seen). Two isolated compact nebulosities are also seen in this hemisphere, $k13$ and
$k14$ -- the latter has an inverted Y-shape, and may represent two (or more) adjoining
elongated knots. Knot $k13$ appears to lie on the periphery of an extended faint structure
($Tube$), described below.

\begin{figure}[htbp]
\vskip 1.0cm
\resizebox{1.0\textwidth}{!}{\includegraphics{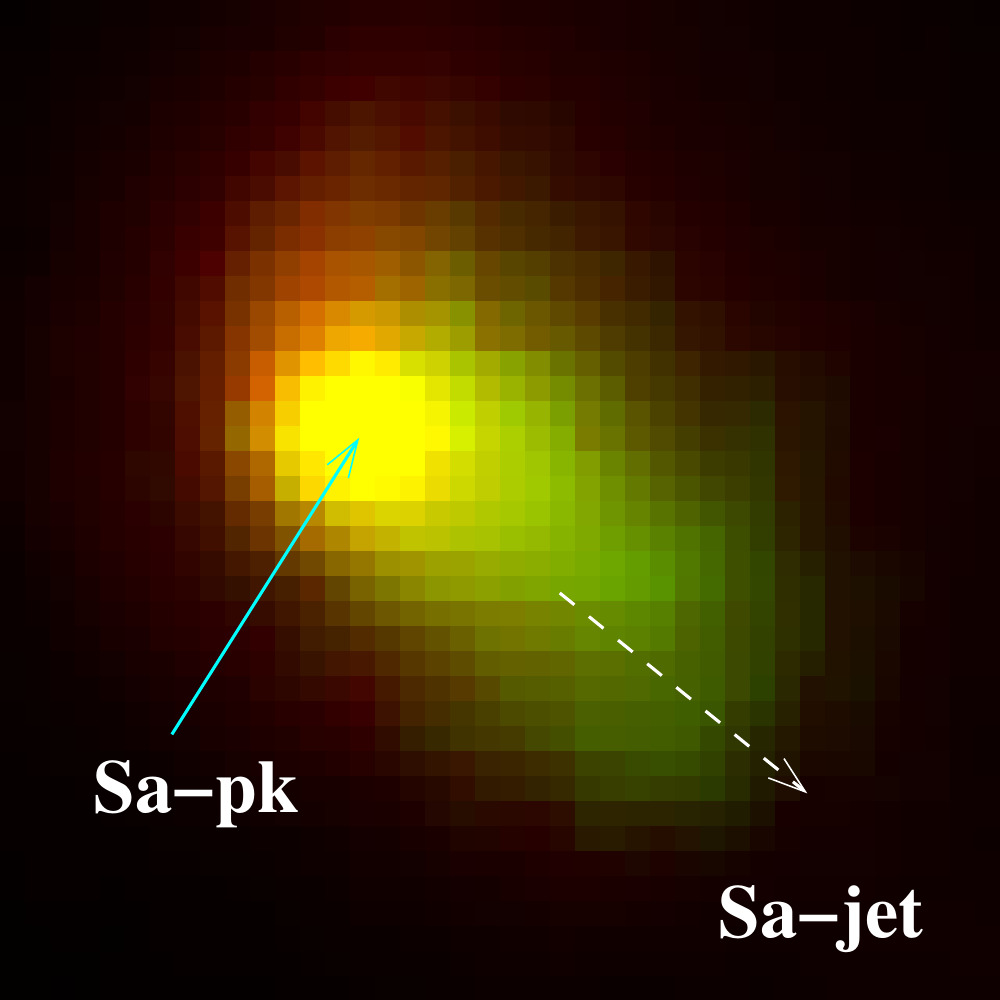}}
\vskip 0.5cm
\caption{HST/ACS color-composite (red: F814W, green: F606W) image of feature {\it Sa},
shown on a linear stretch. The size of the box is 0\farcs5$\times$0\farcs5. The orientation 
of this image is the same in Fig.\,\ref{hst606}}.
\label{hstcen}
\end{figure}

A second compact source ({\it Sb}), comprised of a close stellar
pair with a separation of 0\farcs12 (inset, panel b), is seen 5\farcs64 to the west
(along PA=$-97.6$\arcdeg) of the main source. {\it Sa} has a cometary structure, with a
bright peak at the northeast end {\it Sa-pk}, and a jet-like extension {\it Sa-jet}
towards the southwest, along PA$\sim$230\arcdeg~(Fig.\ref{hstcen}), which is nearly
perpendicular to the average direction of the knot spray seen in the F606W image. 

\begin{figure}[htbp]
\vskip 1.0cm
\resizebox{1.0\textwidth}{!}{\includegraphics{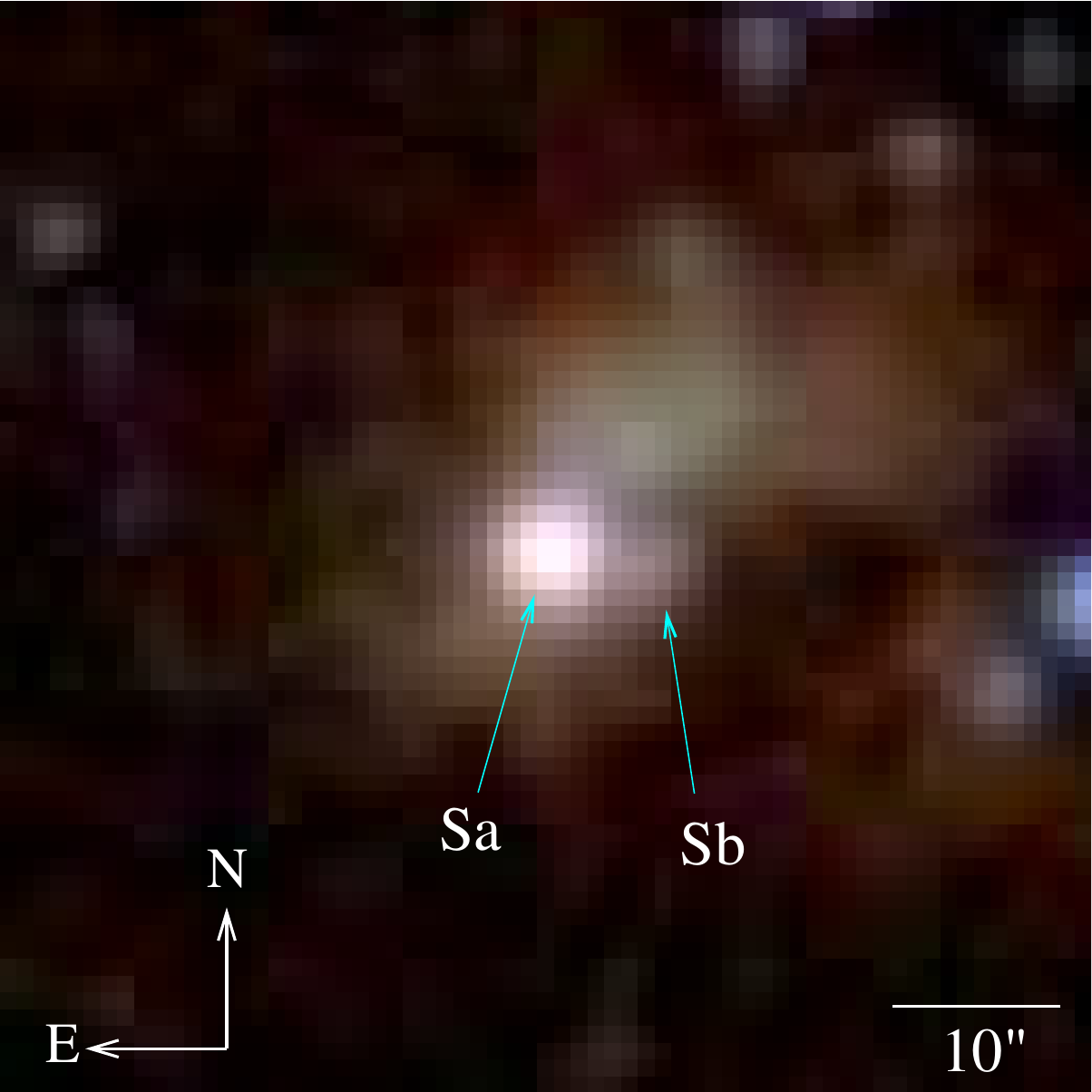}}
\vskip 0.5cm
\caption{Extended nebulosity in the vicinity of IRAS\,05506 is seen in this
color-composite 
image made using 
2MASS $J$ (blue), $H$ (green) \& $K_s$ (red) band images. Each individual image 
is displayed on a {\it log stretch}). Compact sources $S_a$ and $S_b$ are marked for
reference.
}
\label{2mass}
\end{figure}

At the location of {\it Sa-pk}, the NICMOS F110W (Fig.\ref{hst606}c), F160W and F205W (not
shown) images show a point source of increasing brightness with wavelength. For this point
source, we derive fluxes of 6.5, 54, and 190 Jy, respectively, at 1.1, 1.6 and
$2.1\micron$~from the F110W, F160W and F205W images, using the photometric calibration
information provided by STScI\footnote{details at {\tt
http://www.stsci.edu/hst/nicmos/performance/photometry/postncs\_keywords.html}}.
The $1.1\micron$~flux is less reliable than the 1.6 and $2.1\micron$~fluxes because the star
is relatively less bright compared to the surrounding nebulosity. The lack of PSF structure
towards {\it Sa-pk} in the optical images indicates that the central star is not being seen
directly at optical wavelengths. Archival 2MASS images of IRAS\,05506 and vicinity
(Fig.\,\ref{2mass}) show that {\it Sa} lies atop extended diffuse nebulosity which can be
easily seen extending as far as $\sim20{''}-25{''}$ towards the northwest.

The position of {\it Sa-pk} as measured from the F606W HST image is (J2000) RA = 05:53:43.55,
Dec =24:14:44.0. The 2MASS All-Sky Catalog of Point Sources (Cutri et al. 2003) shows only one
source (05534356+2414447\footnote{the 2MASS source names encode the J2000 coordinates of the
sources}) within a $\sim4{''}$ radius of {\it Sa}, and since the uncertainty in the absolute
HST astrometry is $\lesssim1{''}$, the latter represents the near-IR counterpart of {\it Sa}.
The next nearest 2MASS source is 05534318+2414441, located $\sim5{''}$ to the west of {\it
Sa}, and represents the near-IR counterpart of {\it Sb}.

\begin{figure}[htbp]
\vskip -0.5cm
\resizebox{0.65\textwidth}{!}{\includegraphics{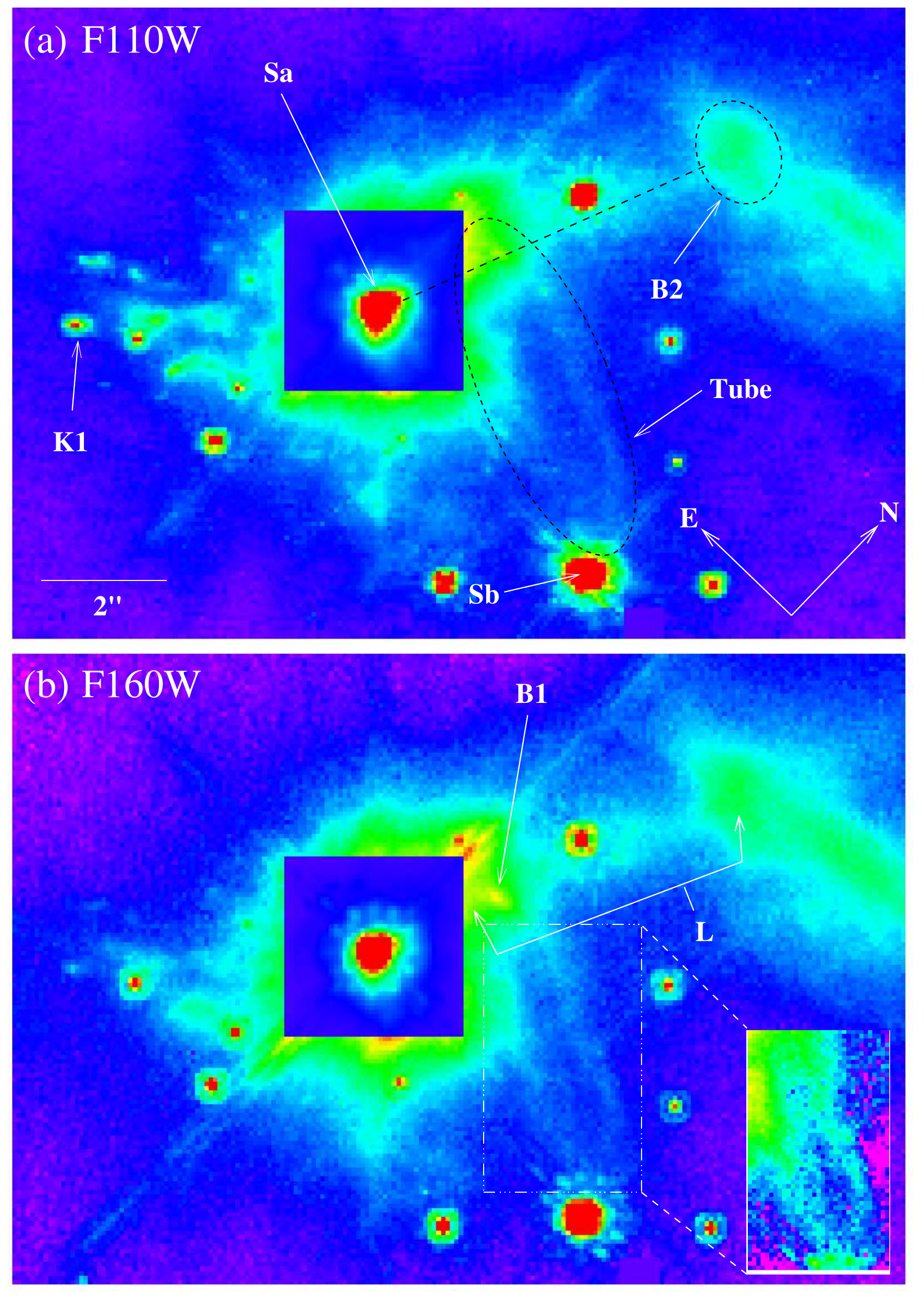}}
\vskip 0.5cm
\caption{(a) F110W and (b) F160W NICMOS images of IRAS\,05506 (shown with a square-root stretch
and false-color intensity scale) highlighting the presence of low-contrast features. The
intensity in the very bright central region surrounding {\it Sa} has been scaled down by a
factor 0.06 (0.03) in the F110W (F160W) image in order to better show the structure of the
surrounding faint nebulosity. The large
ellipse (panel a) encircles a tube-like structure ($Tube$); inset in panel b (lower right
corner) covers a $2.5{''}\times4.3{''}$ patch around the $Tube$ with a different stretch to
show its faint parts more prominently (inset shows $I^{0.25}$, where $I$ is the intensity).
The dashed straight line (panel a)
marks a roughly linear structure, $L$, joining two locally bright regions ($B1$ and $B2$). The 
small ellipse in panel a shows the location of $B2$, which is more diffuse and extended
compared to $B1$. The spiky linear structures oriented at $\pm$45\arcdeg~to the horizontal
which appear to emanate from {\it Sa}, including the 3 bright yellow/orange streaks seen to
the east of $B1$, are diffraction spikes produced by the telescope optics. Note that the
orientation of these images is different from the one in
Fig.\,\ref{hst606}. 
}
\label{nic2}
\end{figure}

We note the presence of two intriguing features in the vicinity of {\it Sa} and {\it Sb}, in
the F110W and F160W NICMOS images (Fig.\,\ref{nic2}). One is a tube-like, limb-brightened
structure (labelled as ``$Tube$" in the figure) that appears to originate from near {\it Sa}
and extends towards {\it Sb}. The second is a roughly linear structure, $L$, joining two
locally bright regions ($B1$ and $B2$); the vector joining these regions passes through {\it
Sa}. $B1$ is bright, relatively compact and appears to be located on the limb brightened
periphery of the $Tube$ feature, whereas $B2$ is much fainter, larger in size and diffuse and
part of an extended nebulosity aligned roughly east-west. In the F205W image, since {\it Sa}
is much brighter relative to the surrounding nebulosity, only those portions of the $Tube$ and
$L$ features which are relatively more distant from {\it Sa} can be clearly seen. The
similarity of the features in the near-IR images indicates that these are dusty structures
seen in scattered ambient starlight. We speculate that the $Tube$ might be a dense-walled
cylinder of
molecular gas and dust and may represent the wake produced by the motion of {\it Sb} through
the dense
interstellar cloud detected in our NH$_3$ mapping. The feature $L$ might also be a wake
produced by the motion of {\it Sa}, or may represent a highly-collimated ejection from {\it
Sa}. Near-infrared emission-line imaging and spectroscopy is needed to better understand the
nature of the $Tube$ and $L$ structures.


\subsection{Optical Spectroscopy}
\label{opt-spec}
\begin{figure}[htbp]
\vskip 1.0cm
\resizebox{1.0\textwidth}{!}{\includegraphics{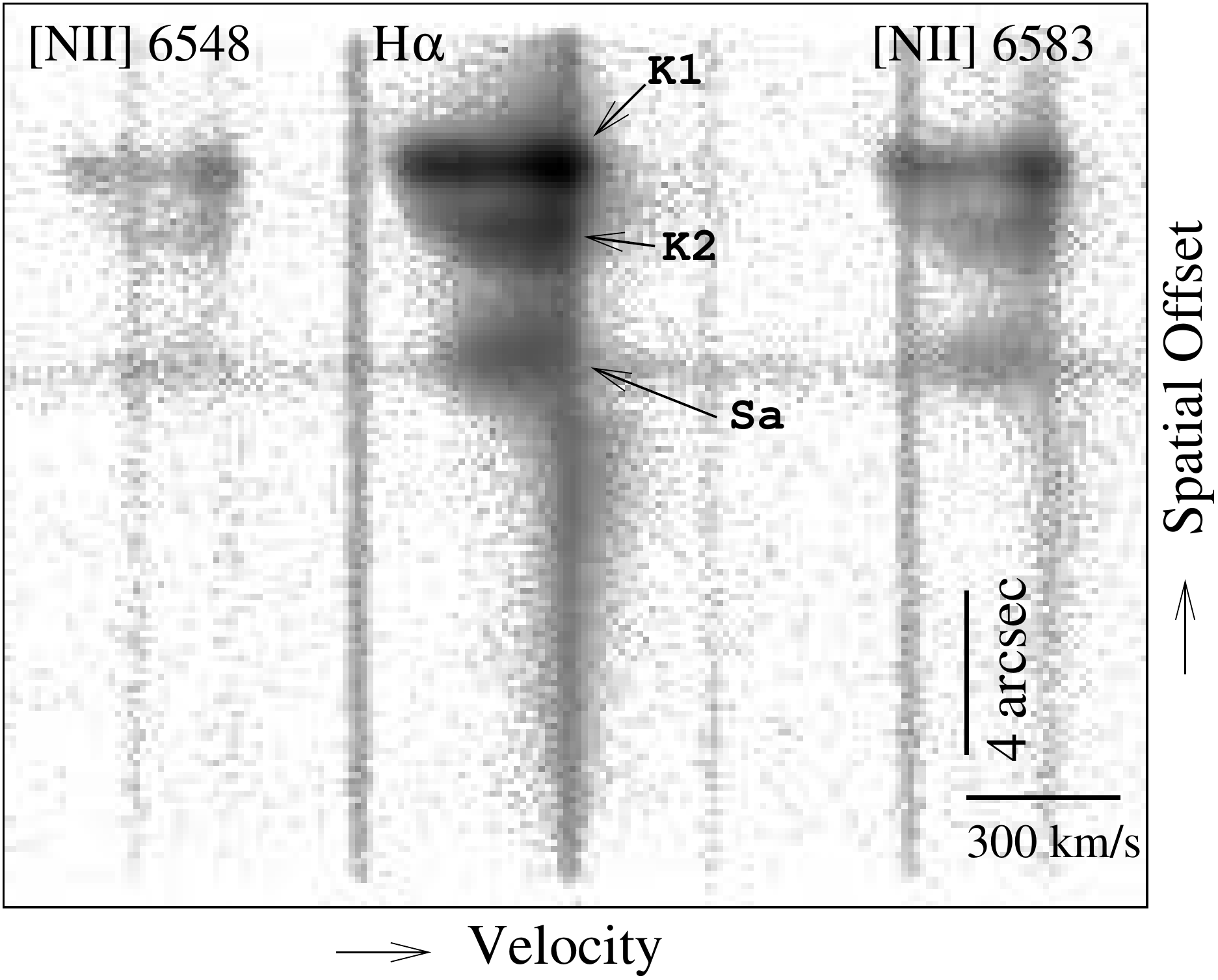}}
\caption{Keck/ESI long-slit spectra of IRAS\,05506 showing high-velocity
outflows and bow-shock structures in the H$\alpha$ and [NII] lines. The narrow vertical
features are due to night-sky emission lines.}
\label{esi}
\end{figure}

The slit was aligned along a vector which passes through the knots $K1$, $K2$ and {\it
Sa}. The
optical spectra of knots $K1$, $K2$ show high-velocity H$\alpha$ and forbidden line
emission such as, e.g., [NII]$\lambda\lambda$6548,\,6583\AA,
[SII]$\lambda\lambda$6716,\,6731\AA, [OI]$\lambda\lambda$6300,\,6364\AA, [Ca
II]$\lambda\lambda$7291.5,7323.9\AA~and many lines of
[FeII]. The position-velocity structure and wide velocity widths ($\sim$350\,\kms) seen
clearly in the strongest of these lines (H$\alpha$, [NII], [SII]) are characteristic of
bow-shock emission (Fig.\ref{esi}) resulting from the interaction of a knotty, high-velocity
outflow (or outflows)  
with dense, ambient material. The knots only show emission lines and no
continuum, similar to Herbig-Haro objects. High-velocity emission is also seen at the
location of {\it Sa}, most likely associated with {\it Sa-jet}. In each case we detect
blue-shifted emission only, which can be interpreted in terms of the near side of a
bipolar outflow in which the far (i.e. red-shifted) side, if present, is hidden from our
view due to local extinction. Unlike the knots, {\it Sa} also shows a very red optical
continuum.

An analysis of the spectrum (Sa\'nchez Contreras et al. 2008) reveals numerous metallic
absorption lines due to Ti I, Fe I, and Ca II in the $\sim8400-8800$\AA~wavelength region
where the continuum from the central star is less extincted and thus detected with much
higher signal-to-noise than at shorter wavelengths. In particular, the Ca II triplet
(8498, 8542, 8662\AA) is quite strong.  The Paschen lines, in the same window, are
noticeably absent - Sa\'nchez Contreras et al. (2008) conclude that the most likely
spectral type of the star is consistent with late G to early K.


Long-slit optical spectroscopy of IRAS\,05506 was reported by Manchado et al. (1990) with
a 1\farcs5-wide slit oriented along PA=80\arcdeg; they found an emission-line spectrum
from an east ``lobe", but only continuum from a west ``lobe"; the lobes being separated
by 5$''$. Our HST images show that, if the ``east" lobe is source {\it Sa} in that study,
as is likely, then the west ``lobe" is source $Sb$.


\subsection{Millimeter-Wave Interferometry}
\begin{figure}[htbp]
\vskip 1.0cm
\resizebox{1.0\textwidth}{!}{\includegraphics{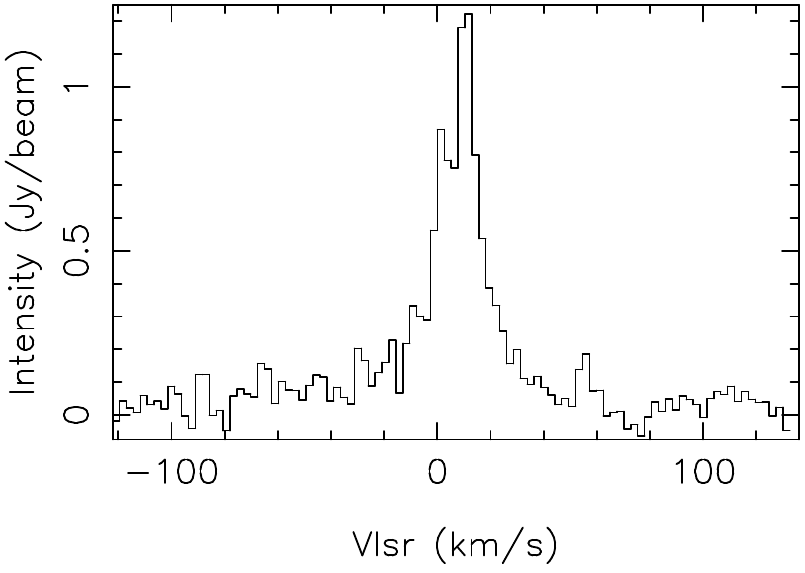}}
\caption{Average intensity spectrum
extracted from a 3${''}\times3{''}$ box centered on the peak intensity position in the
OVRO CO J=1--0 map.}
\label{ovro-cen}
\end{figure}

\begin{figure}[htbp]
\vskip -2.0cm
\resizebox{1.0\textwidth}{!}{\includegraphics{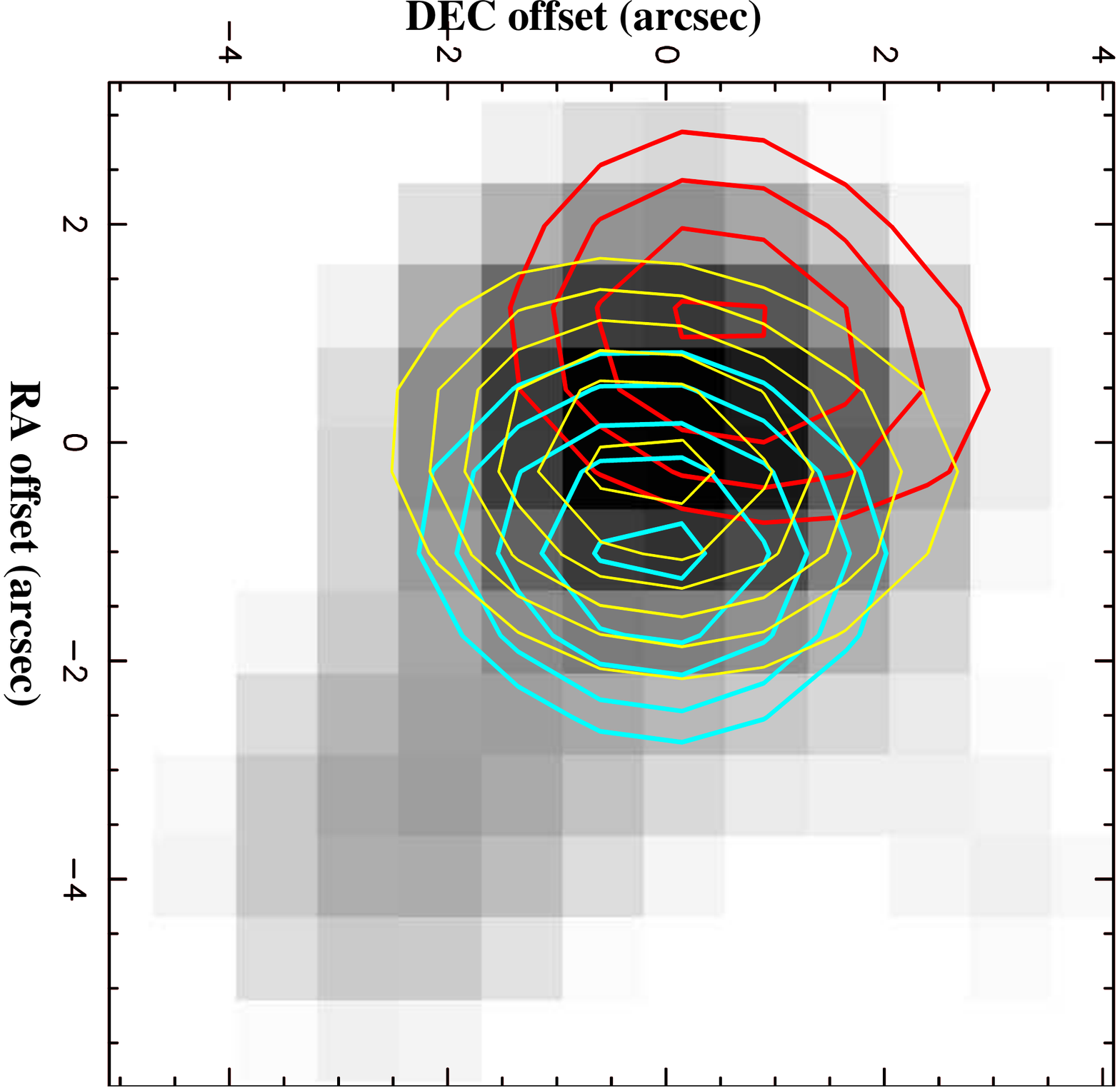}}
\vskip 0.5cm
\caption{OVRO CO J=1--0 (contours) and 2.6\,mm continuum (grey-scale) map of IRAS\,05506
($S_a$). The clean beam for the CO (continuum) map has a FWHM of $3\farcs9 \times
2\farcs9$, PA $-24^{\circ}$ ( $4\farcs6 \times 3\farcs2$, PA $-15^{\circ}$). The yellow,
red and cyan contours show the CO intensity integrated over the $V_{lsr}$ velocity ranges
[2.6,10.4], [49.4:119.6], and [-36.4:-106.6]\,\kms, respectively. The corresponding values
of the peak contour intensity and (spacing) are 6.3 (0.66), 3.8 (0.65), and 4.8 (0.65)
(Jy/beam)\,\kms.}
\label{ovro}
\end{figure}

The CO line profile 
is centered at $V_{lsr}=7.6$\,\kms, and has a core with a width (FWHM) of
$\sim$15\,\kms~and weak, broad wings which extend at $>50$\,\kms~on either side
of the line center 
(Fig.\,\ref{ovro-cen}). The core emission is compact ($\lesssim$ 5\arcsec in size)
(Fig.\,\ref{ovro}). A 10\,mJy continuum source (at $\lambda=2.6$\,mm) with an unresolved
core and a faint extension along PA$\sim$230\arcdeg, consistent with the PA of the
optical jet feature, {\it Sa-jet}, is also seen at the same location. A plot of the 
emission in the extreme red-\,($V_{lsr}$ 49.4 to 119.6\,\kms) and 
blue-\,($V_{lsr} -36.4$ to $-106.6$\,\kms) wings (Fig.\,\ref{ovro}) shows that the outflow
is directed roughly along the same PA as {\it Sa-jet} and the extension in the continuum
source. This high-velocity bipolar outflow is thus directed along an axis roughly
orthogonal to the average direction of high-velocity ejections represented by the optical
knot spray (\S\,\ref{opt-spec}).


\subsection{Ammonia, Water, Methanol and Radio Continuum Observations}
With the VLA, we detected a single, spatially-unresolved H$_2$O emission
feature (across three spectral channels) at (J2000) RA = 05:53:43.54, Dec =
24:14:45.2, at an LSR velocity of 5.9\,\kms~and with a peak flux density of 145 mJy
beam$^{-1}$ (small circle in Fig\,\ref{hst606}a). The 1$\sigma$ error in the H$_2$O maser
position is $\lesssim0.1{''}$. Given the astrometric accuracy of the 2MASS catalog
($\lesssim0.1{''}$), we conclude that there is a small, but significant separation
between the locations of the H$_2$O maser peak and {\it Sa} of $0.57{''}$. This small
separation, and the similarity in the H$_2$O maser radial velocity and the systemic
velocity strongly suggests that latter is associated with {\it Sa}.

\begin{figure}[htbp]
\resizebox{1.0\textwidth}{!}{\includegraphics{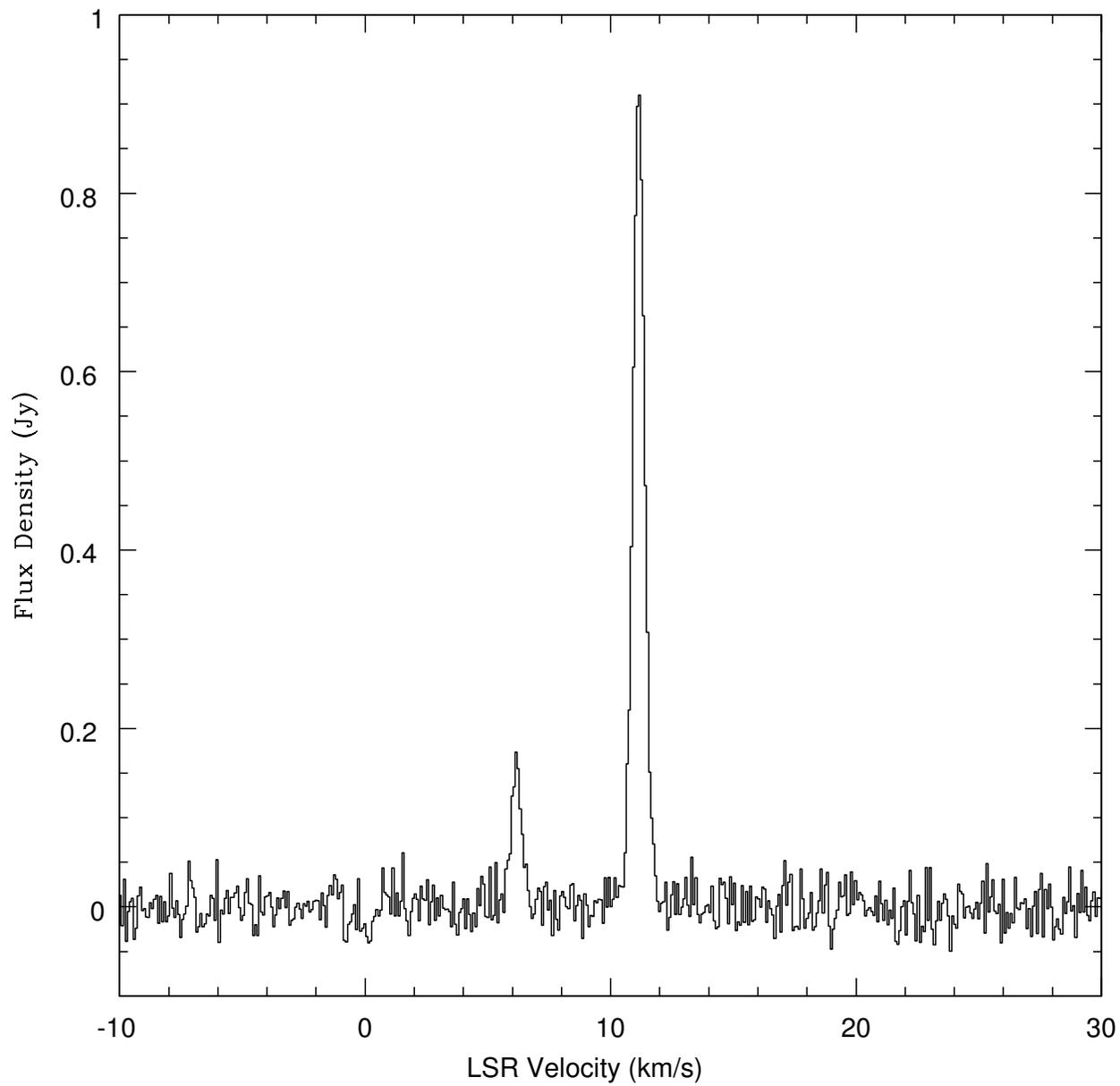}}
\vskip -1.5cm
\caption{Peak H$_2$O emission spectrum of IRAS\,05506 observed with the GBT.}
\label{h2o-gbt}
\end{figure}

With the GBT, we detected emission in the H$_2$O and the NH$_3$\,(1,1) and (2,2) lines. In
H$_2$O, two velocity features are detected; one at 6.16\,\kms~and another at
$V_{LSR}=11.15$\,\kms~(Fig.\,\ref{h2o-gbt}). The FWHM widths (uncertainties) of these
features, derived from Gaussian fits are 0.51 (0.04) and 0.56 (0.01)\,\kms. Both velocity
features are spatially unresolved. These, as well as the NH$_{3}$ 1,1 brightness 
distribution, 
peak at the same raster position, which lies $\sim$8\farcs4 northwest of the VLA H$_2$O
maser position. But given the pointing uncertainties of the GBT and its large beamwidth,
this offset is not significant. The H$_2$O 6.16\,\kms~feature is most likely the same as
that detected with the VLA (hereafter 6\,\kms~feature). The integrated fluxes (errors) of
the H$_2$O maser features for the 2 epochs, June 14, 2005 and Feb 28, 2006, are as
follows: 0.163 (0.017)\,Jy\,\kms~and 1.03 (0.056) Jy\,\kms~for the 6\,\kms~feature and
$<$0.051\,Jy\,\kms~(3$\sigma$) and 12.25 (0.075)\,Jy\,\kms~for the 11\,\kms~feature.
Thus, over the 1.3\,yr period between the two epochs, both features increased
dramatically in strength (the 11\,\kms~feature by a much larger factor), indicating that they
are physically associated and/or the variations represent a common
response to the same stimulus (e.g. passage of a shock wave).

\begin{figure}[htbp]
\resizebox{1.0\textwidth}{!}{\includegraphics{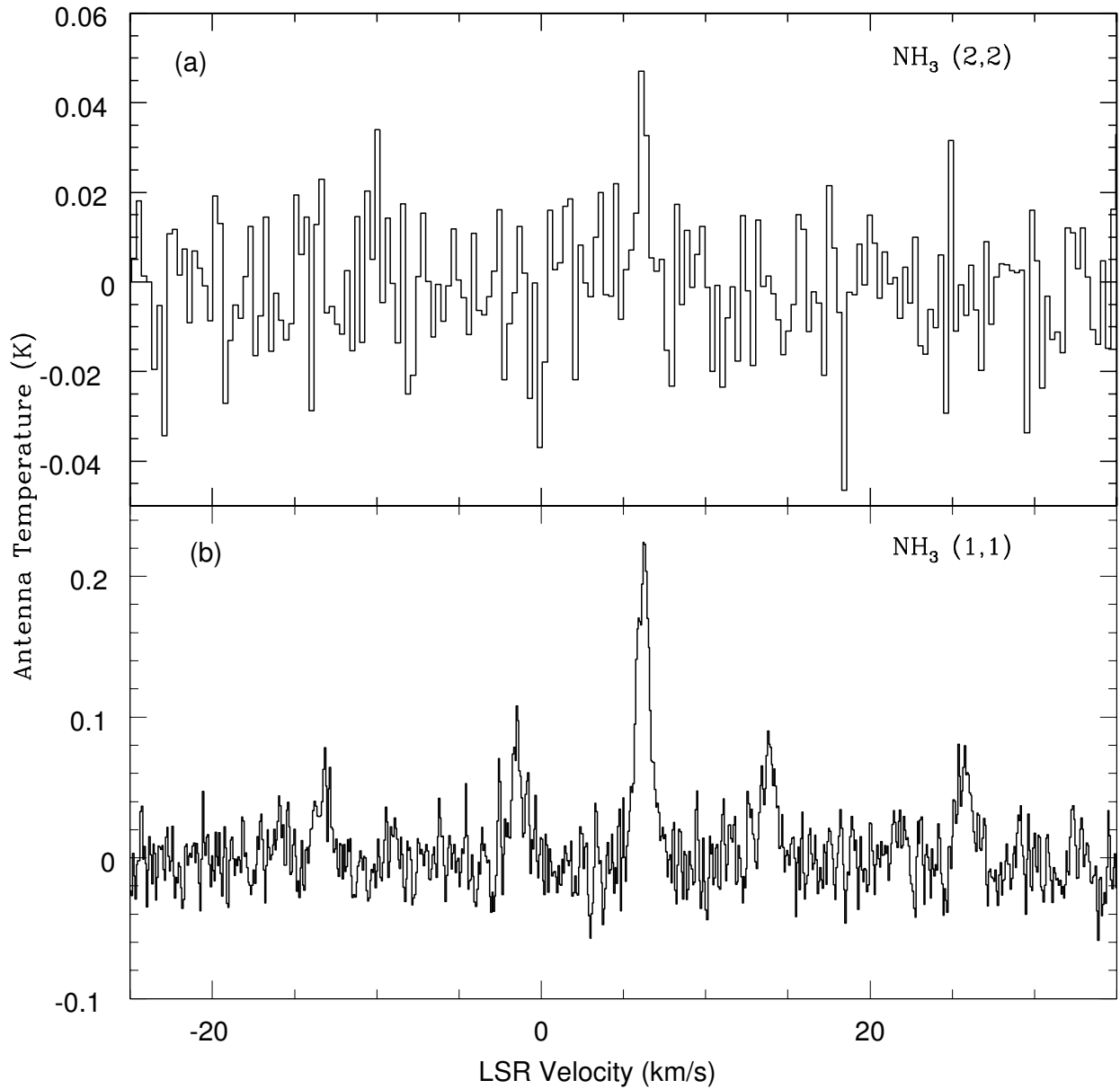}}
\vskip -1.5cm
\caption{The ammonia emission observed with the GBT at the peak raster
position.  (a) The (2,2) transition, boxcar smoothed by a factor
4 (to a resolution of 0.32\,\kms), and resampled by the same
factor.  (b) The (1,1) transition, without smoothing (resolution
of 0.08\,\kms).}
\label{nh3-gbt}
\end{figure}

The NH$_{3}$ lines are surprisingly narrow: the FWHM of the higher signal-to-noise 1,1
line is 0.96\,\kms~(discussed further in \S\ref{discus}). The NH$_{3}$
emission line profile 
in the main hyperfine component is centered at $V_{LSR}=6.21$\,\kms (Fig.\,\ref{nh3-gbt}). The
NH$_3$ emission is
extended towards the east and northwest; the region of emission is likely slightly larger
than 1$'$ (the raster was not large enough to measure the full extent of the ammonia
emission). We detect weak ammonia (2,2) emission (only in the main hyperfine component,
and after smoothing the spectrum by a factor of four), at the position of the peak of the
(1,1) emission. No methanol lines were detected.

Our 8\,GHz continuum observations show no continuum emission (3$\sigma$
upper limit of 150 $\mu$Jy) within a field of approximately 40\arcsec~diameter centered on
the position of IRAS\,05506. From the 1.4\,GHz NVSS all sky survey data, we obtain an
upper limit of 1200 $\mu$Jy.  



\section{Physical Properties of IRAS\,05506}
The steep increase in {\it Sa-pk}'s flux from 0.6 to 2\micron~strongly indicates that it
is a highly extincted stellar source which most likely is driving the
energetic outflow responsible for the whole spray of knots. Consistent with this
interpretation, we find a bright source in the MSX (Midcourse Space Experiment: Egan et
al. 1999) catalog located at the position of {\it Sa-pk}, i.e., at
(J2000)\,RA=05:53:43.55, DEC=24:14:44.0. In this section, we derive the physical
properties of the central source and its environment, many of which depend on the
distance to the source. We estimate a formal kinematic distance of 2.8\,kpc to
IRAS\,05506, using the formulation given by Goodrich (1991). We have taken the average of
the source radial velocities from our ammonia and CO data, i.e., 6.9\,\kms~as the
systemic velocity, and assumed Galactic parameters $R_0=8$\,kpc, $\Theta=220$\,\kms~and
circular rotation. But given the galactic longitude of l$_{II}=185$\arcdeg, and allowing
a typical molecular cloud velocity dispersion of $\sim$10\,\kms, the possible distance
range is not well constrained. In this paper, we adopt a fiducial distance, D=2.8\,kpc,
and scale all of our estimates with D. At this distance, the bolometric luminosity of
IRAS\,05506 derived from integrating over its (very red) SED, is, $L\sim5100$\ls. The
integration spans the full range of wavelengths covered by ISO archival spectra of this
object, i.e., $2.4-198$\micron. The uncertainty in estimating $L$ due to flux beyond the
wavelength limits of the SED is $<$5\%. 

Given the spectral type (late G -- early K, i.e. T$_{eff}\sim5000-4400$\,K) of the central
star inferred from our optical spectroscopy, and the above luminosity, we can locate 
IRAS\,05506 on the HR diagram and compare with evolutionary models. Inspection of the
pre-main-sequence stellar evolutionary models of
Bernasconi \& Maeder (1996) shows that the above values of T$_{eff}$ and $L$ lie between
the tracks for stars with main-sequence masses of 9 and 15\ms, and at the low temperature
end of these tracks -- thus close to the epoch where deuterium-burning is initiated.

\subsection{The Optical Knot Spray: Interstellar Bullets}
We propose that the knots are Herbig-Haro objects based on their forbidden-line emission; they
are seen most prominently in the F606W
image compared to the images at longer wavelengths because this filter covers the
H$\alpha$ and [NII] lines.
The [SII]$\lambda\,6716/\lambda\,6731$ ratio, $R_{[SII]}$, is about 0.6 (0.5) for the low
(high)-velocity emission in the knots, implying electron densities of about 3300 (8000)
cm$^{-3}$
assuming an electron temperature of $10^4$ K (Canto et al. 1980). Using the average value
of $R_{[SII]}$ for knot K1 (0.54), we derive $n_e=6000$ cm$^{-3}$, and approximating
$K1$'s shape with a prolate ellipsoid having major (minor) axis=0\farcs29 (0\farcs15), we
find, assuming $n_e=n_H$, that its mass is, $M_{K1}=2\times10^{-5}$(D/2.8\,kpc)$^3$\ms.
The knots $K2-K4$ are somewhat fainter (by factors of 2-3) than $K1$, and therefore may
have correspondingly lower, but comparable, masses.

The morphology and kinematics of the fan-like spray of elongated knots in IRAS\,05506 are
similar to the
interstellar bullets seen in H$_2$(2.12\micron) and [FeII](1.65\micron) emission towards
the Orion molecular cloud (e.g., Allen \& Burton 1993). Allen \& Burton
(1993) derive a minimum mass of $\sim\,10^{-5}$\ms~for the Orion bullets, comparable to
our value for knot $K1$. The opening angle of the bullet spray in IRAS\,05506 is about
45\arcdeg~(70\arcdeg), measured from knot $K8$ to $k10$ ($k12$), hence somewhat smaller than
the almost 90\arcdeg~opening angle covered by the Peak 1
region (located northwest of BN and Source I) shown in Nissen et al. (2007). The Orion bullets
are typically
$2{''}-4{''}$ in angular extent (Tedds, Brand \& Burton 1999); if we assume that the
physical
size of the IRAS\,05506 bullets is similar, then their smaller angular sizes suggest that
IRAS\,05506 is about a factor 10 more distant than Orion, i.e., 4.5\,kpc. The total
(angular) radial extent covered by the bullets in the Peak 1 Orion region is about
$35{''}-40{''}$, i.e., a factor 6-7 times larger than that measured for the IRAS\,05506
bullets region -- consistent with a larger distance to the latter by a similar factor, if
we assume similar physical extents for the bullet regions of these two sources. The 6--10 times
larger distance to IRAS\,05506 could also account naturally for
the fact that the number of high-velocity knots in IRAS\,05506 is much smaller
than that found toward Orion, since we presumably detect only the most extreme (i.e., brightest
and largest) members of a
larger collection of knots. Similarly, it may be that the opening angle of the
IRAS\,05506 ``bullet" region is intrinsically comparable to that found for Orion, but our
current observations lack the sensitivity needed to detect lower brightness knots at wider
opening angles because of IRAS\,05506's much larger distance. Furthermore, if the filamentary
structures seen in the F110W image (Fig.\,\ref{jband2}), together with knots $k13$ and $k14$
represent the unresolved or partially resolved counterpart to the bullet spray seen to the
east of {\it Sa}, then the similarity to the Orion bullet spray becomes even more pronounced. 
For knot $K1$, which is the most
distant knot from {\it Sa}, we estimate an expansion time-scale of 200\,(D/2.8\,kpc)\,yr ,
under the assumption that the FWZI of the velocity-profile of a
bow-shock is equal to the intrinsic speed of the bullet producing the bow shock (Hartigan,
Raymond \& Hartmann 1987). For knot $K2$, which is closer to {\it Sa} than $K1$, the expansion
speed, inferred as above, is smaller ($\sim$300\,\kms), and we find an expansion time-scale of
about 170\,(D/2.8\,kpc)\,yr. We don't
know if all the knots were ejected quasi-simultaneously over some modest range of velocities
from the central source, or result from an episodic outflow which ejects high-speed blobs
intermittently over a wide range of position angles.

In addition, we have examined the literature on a few other examples of possibly
similar systems with high-velocity knots. For example, the
high-mass YSO IRAS\,18151-1208 contains knots spread along two, roughly orthogonal, axes, and
are related to two well-defined collimated outflows
(Davis et al. 2004).  The outflows HH\,212 and HH\,211, which arise from low-mass 
protostars, also have very highly collimated outflows (Lee et al. 2006, 2007).  Thus
these three YSOs are morphologically quite dissimilar to
the knotty ejections from IRAS\,05506, which have an opening
angle of at least 44\arcdeg. The high-mass YSO 
IRAS\,05358+3843, displays a fan of knots seen in near-IR H$_2$
emission, and for this source, Kumar et al. (2002) make a similar analogy to
the Orion bullets.


The optical jet, {\it Sa-jet}, is much bluer than the central source, {\it Sa-pk},
presumably because like the knots, it is dominated by optical line emission. The rather
precise alignment of the position angle of this jet, which is seen out to a radius of
0\farcs25 from {\it Sa-pk}, to that of the much more elongated (6$''$, giving a
deconvolved length of about 4\farcs5) feature seen in the millimeter continuum image, is
remarkable. Is this outflow different in nature than that which produced the bright knots
$K1-K4$, since no such knots appear to be associated with it? It may be that such knots
are present, but not visible due to local extinction. The average large-scale (i.e., over
a 0.5$'$ area) extinction derived from our NH$_3$ data (see \S\,\ref{dust_phys} below) is
$A_V\sim3$, and it is likely that the extinction increases steeply towards {\it Sa}.
Indeed, we find that the line-of-sight extinction to {\it Sa} is $A_V\sim25$ by
comparing each of our observed 1.6 and 2.2\micron~fluxes of the central star with the
values expected based on its bolometric luminosity and effective temperature. We have
used the tabulated average extinction curve in Whittet (1992) to derive the visual
extinction from our near-IR values. A faint patch of nebulosity is seen
in the F606W image at an offset of about 3\farcs5 along the PA of the optical jet, but it
does not have a well delineated structure like the $K1-K4$ knots and is similar to other
faint diffuse nebulosities seen in the vicinity of {\it Sa}, so it is not clear whether or
not it is directly associated with the jet.

\subsection{The Gas and Dust Environment}
\label{dust_phys}
The physical properties of the molecular gas associated with, and/or in the vicinity of
IRAS\,05506 on an arc-minute angular scale, can be inferred from the ammonia data. Using
Gaussian fits to the ammonia line profiles to estimate the peak intensities, and the
formalism given by Wilson, Gaume, \& Johnston (1993), Ungerechts, Walmsley, \& Winnewiser
(1986), and Ho \& Townes (1983), we estimate the opacity of the (1,1) transition
($\tau\sim 0.7$ for the main line) and the rotational temperature between the (2,2) and
(1,1) transitions (T$_{rot} = 12.5\pm2.2$ K). The rotational temperature is a lower limit
to the kinetic temperature, T$_{kin}$; for these low values of T$_{rot}$, T$_{kin}
\approx $T$_{rot}$ (Morris et al. 1973, Danby et al. 1988). Assuming a beam filling
factor of unity, the ammonia column density toward the peak of the emission is
N$_{NH_3}\sim1.1\times10^{14}$cm$^{-2}$. Estimating the region of ammonia emission to
have a diameter of 1$'$, and adopting a fractional NH$_3$/H$_2$ abundance ratio (by
number)\footnote{Pillai et al. (2006) find values in the range (0.7-10)$\times10^{-8}$},
$f(NH_3)=4\times10^{-8}$, we obtain a molecular hydrogen mass in this region of
11\,(D/2.8\,kpc)$^2$\ms. In this calculation, we have accounted for the lower emission at
the outer regions of the source, by assuming that the derived mass is a factor 2 lower
than the value that would result from assuming uniform emission within the 1$'$ region. We
estimate that the large-scale visual extinction towards the center of the cloud is,
$A_V\sim3$, using a ratio $N(H_2)/A_V=0.94\times10^{21}$cm$^{-2}$ mag$^{-1}$ (Frerking,
Langer \& Wilson 1982). 
 
The virial mass of the cloud (assuming an r$^{-1}$ density distribution) is,
$M_{vir}$ = 188(R/pc) ($\Delta$V$_{1/2}$/\kms)$^2$ = 73(D/2.8\,kpc)($\theta$/0.5\arcmin)\,\ms,
where $\theta$ is the cloud radius in angular units, and $\Delta$V$_{1/2}$=0.96\,\kms.
For an r$^{-2}$ density cloud, we get $M_{vir}$=48(D/2.8\,kpc)($\theta$/0.5\arcmin)\,\ms.
The factor 4 (or more) discrepancy between the virial cloud mass and that derived from
the ammonia data may easily be attributed to the large uncertainties in our assumed value
of $f(NH_3)$ and source distance, and/or to non-gravitational linewidths, such as
unresolved outflow motion.

We have very recently (Nov 2007) mapped the $^{12}$CO J=1--0 emission in a $36'\times18'$ 
($\Delta\alpha\times\Delta\delta$) region around IRAS\,05506 using the ARO 12-m
millimeter-wave telescope, and find the presence of a molecular cloud which extends over
at least 30$'$ East-West and 12$'$ North-South. Observations of the $^{13}$CO and
C$^{18}$O J=1--0 emission towards the central region show that the $^{12}$CO line is
optically thick, both in the line core and wings. We will discuss these data in detail in
a future publication; our preliminary analysis shows that the extended cloud which we
detect in $^{12}$CO J=1--0 emission is easily many hundreds of solar masses. 

\begin{figure}[htbp]
\resizebox{0.9\textwidth}{!}{\includegraphics{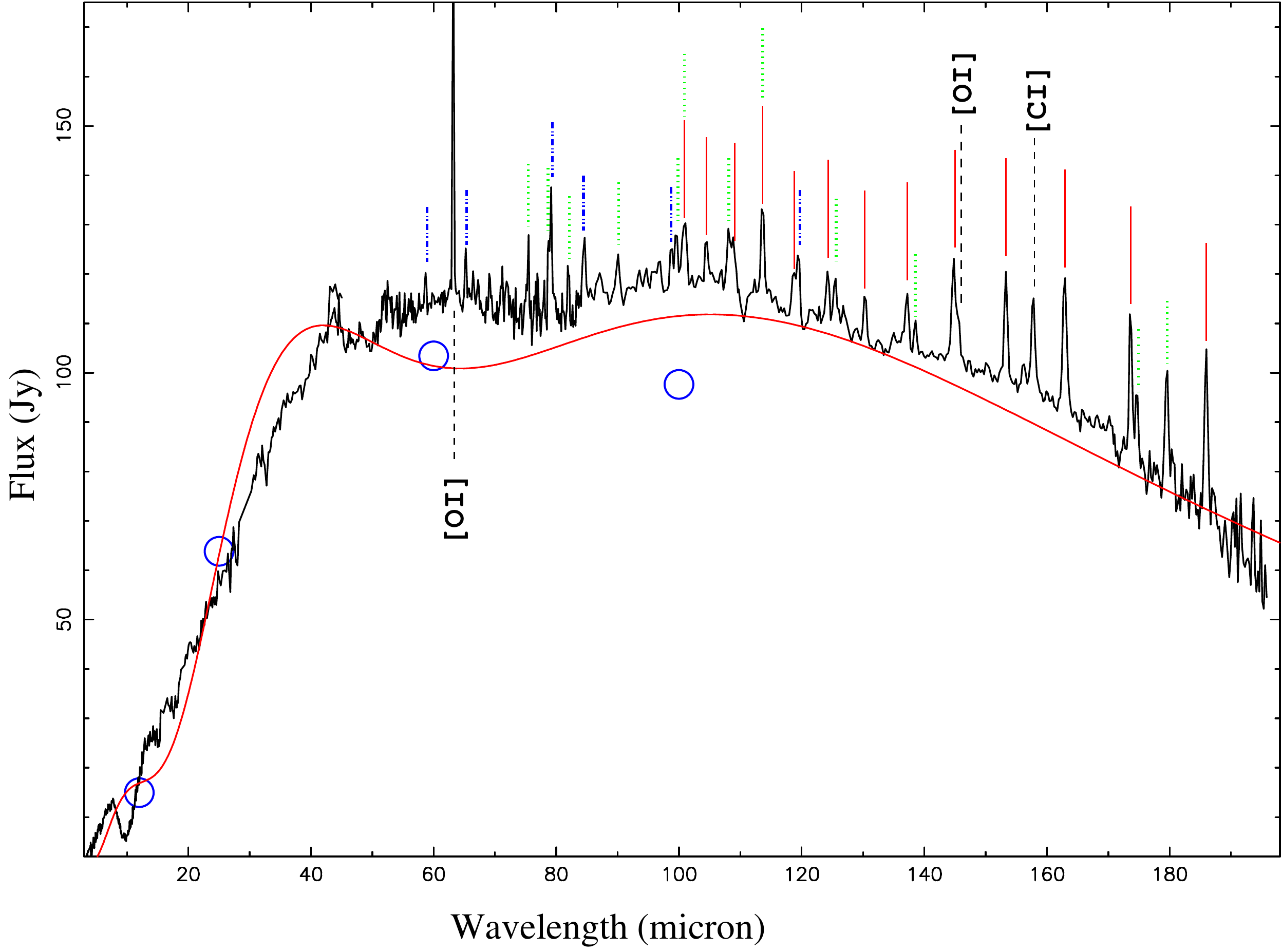}}
\caption{ISO SWS and LWS spectra of IRAS\,05506, showing a deep 10\,\micron~amorphous
silicate absorption feature, and emission in the OI (at 63 and 146\,\micron; dashed vertical
bars) and CII (158\,\micron: dashed vertical bar)
lines as well as numerous far-infrared CO, 
OH and H$_2$O lines. The locations of the CO lines from J=14--13 (186.0\micron) to J=26--25
(100.5\micron) 
are shown by short (solid) vertical red bars, and those of the most prominent lines of H$_2$O
(both ortho- and para-) and OH are marked by green (dotted) and blue (dash-dot-dot) vertical
bars, resprectively. The symbols show broad-band photometric
data from IRAS. The red curve shows a simple dust model fitted to the data.}
\label{iso}
\end{figure}

The molecular cloud detected in NH$_{3}$ emission is presumably associated with the
extended diffuse nebulosity which is seen in the 2MASS images. The nebulosity is most
likely due to scattered light from dust. In order to estimate the dust mass associated
with this cloud, we have retrieved and re-calibrated archival ISO SWS and LWS spectra of
IRAS\,05506 covering the 3-200\,\micron~wavelength range (Fig.\ref{iso}). We find that it
shows a very red continuum, a deep 10\,\micron~amorphous silicate absorption feature, and
emission in numerous far-infrared CO, OH and H$_2$O lines, as well as the OI (at 63 and
146\,\micron) and CII (158\,\micron) lines (Sahai \& Sarkar 2008, in prep). The far-IR
line spectrum of IRAS\,05506 is very similar to that seen towards the low-mass protostar
IRAS\,16293-2422 (Ceccarelli et al. 1998). A multi-component dust model (i.e., with
``cool", ``warm", and ``hot" components, Sahai et al. 1991) assuming a power-law
($\lambda^{-p}$, with $p=1.5$) dust emissivity with a value of 150 cm$^2$g$^{-1}$ (per
unit {\it dust} mass) at 60\,$\mu$m (Jura 1986), provides a modest fit to the SED, and
demonstrates that the bulk of the mass is in cool dust. We find that the cool component
has 0.26\,\ms~of dust at 27\,K, the warm component has $0.22\times10^{-2}$\ms~of dust at
79\,K; both masses scale as (D/2.8\,kpc)$^2$. These imply a total mass of 26\,\ms
(assuming a gas-to-dust ratio, $g_d=100$), which compares reasonably well with that
derived from our ammonia observations. Note that the above modelling can provide only a 
rough estimate of the mass because the dust is expected to be distributed 
over a range of temperatures, depending on the distance from the heating source(s). Sahai et
al. (2007) have compared the results of such modelling with those from more accurate modelling
(using a dust radiative transfer code) for pre-planetary nebulae, and find that the simple
modelling gives dust masses that are lower by factors of 0.6--0.8.


For the mm-continuum source associated with {\it Sa}, we derive a dust mass of
(0.022--0.064)\,(D/2.8\,kpc)$^2$\ms, assuming 79--27\,K for the temperature of the dust,
and that $p=1.5$ is valid over the $\lambda=60$\,\micron~to 2.6\,mm wavelength range.
Thus the total mass associated with {\it Sa} within a $\sim$4$''$-size region is
2.2--6.4\,\ms (assuming $g_d$ as above). The total molecular mass derived from the CO
flux detected with the OVRO beam, i.e., within a $\sim$3\farcs4-size region, is
0.26\,\ms~assuming a fractional CO/H$_2$ abundance ratio, $f(CO)=10^{-4}$, and taking
T$_{rot}$=27\,K\footnote{if the excitation temperature is higher, the derived mass will
be larger; e.g., for T$_{rot}$=79\,K, the derived mass is a factor 2.5 larger}. Because
the CO is optically-thick, this value represents a lower limit.

The inadequacies of the SED fit from our simple model indicate that the source has dust
over a wide range of temperatures. Sahai \& Sarkar (2007) have therefore used the online
tool provided by Robitaille et al. (2007), which provides least-squares fits of
pre-computed disk-envelope models of young stellar objects (YSOs) to user-defined SEDs, in
order to fit IRAS\,05506's SED.
Given an input distance range of 0.1--5\,kpc, the fitting tool finds that the first three
best-fit models to the SED of IRAS\,05506 have, respectively, distances of D=2.2, 2.6,
3.0\,kpc (implying bolometric luminosities in the range (3150-5850)\,\ls), central star
masses of M=8.5, 10.6, 11.2\ms, and T$_{eff}$=4190, 4430, 4280\,K.
These T$_{eff}$ values are consistent with the spectral type of the central star inferred
from our optical spectroscopy.

\subsection{Molecular Outflow: Mass and Dynamics}
We estimate the mass ($M_{bip}$), mass-outflow rate (\mdot$_{bip}$), scalar momentum
($P_{sc}$), mechanical energy ($E_{mech}$) and the mechanical power ($L_{mech}$) in the
high-velocity bipolar outflow using the integrated CO J=1--0 line profile from our OVRO
data. The derived values for $M_{bip}$, $P_{sc}$, and $E_{mech}$ scale as
(D/2.8\,kpc)$^2$, whereas \mdot$_{bip}$ and $L_{mech}$ scale as D/2.8\,kpc. The scalar
momentum and the mechanical energy are computed using the methodology described in
Bujarrabal et al. (2001). As before, we assume T$_{rot}$=27\,K, $f(CO)=10^{-4}$, and the
emission to be optically thin -- since the
CO emission is optically thick, our derived values are lower limits. In this analysis, we
use the J=1--0 emission flux over the $V_{lsr}$ velocity ranges [$10.4:119.6$]\,\kms~for
the red component and [$2.6:-106.6$]\,\kms~for the blue component of the outflow shown in
Fig.\,\ref{ovro}. We have conservatively excluded emission in the central three 1\,MHz
channels which cover the line core. We infer an expansion time-scale for the outflow of
185\,yr, by dividing the linear separation between the red and blue outflow components by
the mean separation velocity, 156\,\kms~(derived from the difference between the average
velocities for the red and blue components in Fig.\,\ref{ovro}) and assuming an
intermediate inclination (45\arcdeg) of the outflow axis relative to the line-of-sight.
We find $M_{bip}>0.18$\ms, \mdot$_{bip}>9.7\times10^{-4}$\my, and
$P_{sc}>8.6\times10^{38}$ g\,cm\,s$^{-1}$ (4.3\,\ms\,\kms). The mechanical energy in the
bipolar outflow is, $E_{mech}>4.6\times10^{45}$ erg; dividing this by the expansion
time-scale, we obtain $L_{mech}>7.9\times10^{35}$ erg s$^{-1}$. Hence the ratio of the
mechanical power to the stellar luminosity is, $L_{mech}/L>0.04$, which is significantly
larger than the values typically found for YSO outflows ($10^{-3}-10^{-2}$; Lada 1985). 
The outflow cannot be driven by radiation pressure: the expansion time-scale,
$\sim$185\,yr, is much smaller than the time required by radiation pressure to accelerate
the observed bipolar outflow to its current speed, $t_{rad}=P_{sc}/(L/c)\sim 4\times10^4$
yr.

\section{Discussion}
\label{discus}
The morphology of IRAS\,05506 does not resemble that of any known PPN. Although its
luminosity (at a distance of 2.8\,kpc) is commensurate with that expected for PPNs, the
circumstellar mass is significantly larger than that typically found in PPNs ($\lesssim$
1\,\ms, e.g. Sahai et al. 2007). 
Zinchenko, Pirogov \& Toriseva (1998) include IRAS\,05506 (labelled by its Galactic
coordinates, as G\,184.96-0.85) in their study of dense molecular cores in regions of
massive star formation, and find CO J=1--0, CS and C$^{34}$S J=2--1 emission towards this
object using the Onsala 20-m telescope (beamsizes 35$''$ and 40$''$). The radial velocity
of the CS line, 6.2\,\kms~$V_{lsr}$, agrees with that found from our NH$_3$ observations,
so there can be little doubt that the CS emission is probing the same molecular cloud
core as NH$_3$. Our recent mapping of CO J=1--0 emission towards IRAS\,05506 with the ARO
12-m dish shows the presence of a very extended molecular cloud with a total mass of at
least several hundred solar masses.

IRAS\,05506 is, in all likelihood, not a PPN. 
The 1612 MHz OH maser line profile in
IRAS\,05506, observed with the Arecibo dish (Chengalur et al. 1993), does not show the
typical double-peaked shape seen towards most AGB stars or PPNs. The main lines of OH at
1665 and 1667\,MHz were detected by Lewis (1997) 
at Arecibo; the emission covers about 15\,\kms~in the 1612 MHz line and
$\sim$40--45\,\kms~in the main lines.
The combined presence of maser (H$_2$O and OH), NH$_3$, CO, CS \& C$^{34}$S and relatively
strong far-IR emission, strongly argues for a dense, star-forming core being associated
with IRAS\,05506. Palagi et al. (1993) also classify
IRAS\,05506 as a star-formation region, on the basis of a principal component analysis of
the IRAS colors of all H$_2$O maser sources in the Arcetri Catalog (Comoretto et al.
1990) which had IRAS PSC counterparts.
The very short expansion time-scale of the optical knots and the fast molecular outflow
indicates that the source {\it Sa} is a very young protostar.

The presence of OH maser emission argues for IRAS\,05506 not being a low-mass YSO, since
there are no known OH masers toward low-mass YSOs. There have been surveys for OH maser
emission from, e.g., T Tauri stars (Chandler et al., private communication), but no
detections have been made and we are unaware of any published detections of OH masers
toward low-mass YSOs. Given our calculated luminosity at D=2.8\,kpc, our distance estimate
would have to be lower by at least a factor 4 for the luminosity to be in the range for
low-mass YSOs.

Is IRAS\,05506 an intermediate-mass or massive YSO? In a search of the literature on OH masers
towards star-forming regions, we found only one region (L1287) which does not appear to be
forming massive stars and which shows an OH maser (Wouterloot, Brand, \& Fiegle 1993). This
region contains the FU Ori-type objects RNO 1B, RNO 1C and an IRAS source (IRAS\,00338+6312)
whose luminosity indicates it to be an intermediate-mass YSO. So there is a small probability
that IRAS\,05506 is associated with an intermediate-mass star (Quanz et al. 2007 and
references therein). Further, the total molecular cloud mass derived from the ammonia data is
rather small for a massive star, because the implied efficiency of star formation would have
to be significantly higher than is usually assumed to be the case. Note, however, that this
issue is a lesser concern if the source were at a distance exceeding 5\,kpc, because of the
scaling of the cloud mass with the square of the distance. Also, the extended molecular cloud
towards IRAS\,05506 seen via CO mapping harbors several hundreds of solar mass of material
(for D=2.8\,kpc) within a region 3 times bigger in size than that probed by our ammonia data.
Although CO masses are not the best measure of the mass of a cloud core, the total mass of the
associated cloud is likely well above what is needed to support massive star formation. It is
perhaps unwarranted to be too concerned about the relatively low value of the cloud core mass,
given the possibility that the core has been strongly altered or dissipated in the star
formation process, especially in the presence of the observed strong outflows.

The multiple optical knots in IRAS\,05506 imply that material has been ejected from the central
source in the form of high-speed bullets rather than in a continuous fashion. The presence of
this wide-angle spray of bullets in IRAS\,05506 indicates, by analogy with Orion, that the
launching source may be a massive protostar. A massive star, depending on its evolutionary
status, could produce an ultra-compact HII region with detectable continuum emission. But
since we do not find a measurable radio continuum towards IRAS\,05506, either it is not a
massive star, or (i) the central HII region is still very small, (ii) the source is
sufficiently distant that its radio continuum is below our noise limit, (iii) the central star
has been caught at a very early age while its temperature is still too low to provide
sufficient UV flux for ionisation -- this possibility is consistent
with the late spectral type of the central star inferred from our optical spectroscopy.

We compare IRAS\,05506 with the high-mass protostellar objects (HMPOs) identified by
Sridharan et al. (2002, hereafter Setal02). These HMPOs are dense molecular cores detected in a
CS J=2--1
survey of massive star-forming regions, which are bright at FIR wavelengths
($F60>90$\,Jy) and generally have low (or absent) centimeter radio fluxes. IRAS\,05506
satisfies these selection criteria for HMPOs, but shows a significantly smaller NH$_3$
line-width ($\Delta$V$_{NH_3}$=0.96\,\kms) and kinetic temperature (12.5\,K) than the
median values for \mbox{HMPOs} ($\Delta$V$_{NH_3}=1.9$\,\kms~and $T_{kin}=18.5$\,K:
derived, as for IRAS\,05506, from single-dish ammonia observations). In fact, $T_{kin}$
for IRAS\,05506 is lower than the smallest value for the 40 HMPOs in which $T_{kin}$ was
measured.

Setal02 have made 2-component dust model fits to the IRAS
fluxes for these HMPOs and find cold dust components with $T_d\sim30-70$\,K, and hot
components with $T_d\sim100-250$\,K, i.e. somewhat higher than the corresponding ones
for IRAS\,05506. The bulk of the luminosity for the HMPOs comes from the cool component,
as we have found for IRAS\,05506. The luminosity of IRAS\,05506 is comparable to those
found for many objects in the HMPO list of Setal02. 
However, the associated IRAS fluxes (at 60\micron; only upper limits to the
100\micron~fluxes are available for most HMPOs) and thus the derived masses of the cool
dust component for HMPOs with comparable (near) kinematic distances are generally higher
than that for IRAS\,05506, by factors of $\sim$1.5 or more. 

IRAS\,05506's ammonia line-width and gas kinetic temperature are even lower than the
median values (1.5\,\kms~and 16.9\,K) for high-mass starless cores (HMSCs: Sridharan et
al. 2005), which are apparently less evolved than HMPOs. HMPOs, but not HMSCs, have MSX
Point Source catalog counterparts, indicative of the presence of a luminous, hot, central
source. IRAS\,05506 may thus be an extreme example of the HMPO class,
lying at the bottom end of the ammonia line-width and kinetic temperature distributions
(see, e.g., Fig.\,3 of Sridharan et al. 2005), and thus at the lower end of the mass
range for these objects. In a plot of HMPO core mass (derived from single-dish mm-continuum
data) versus FIR luminosities (Fig. 6 of Setal02), there are two sources with core masses 
comparable to
that derived from the mm continuum for IRAS\,05506. Given that the latter is based on
interferometric observations, which resolve out flux on scales more extended than about
$4{''}$, whereas the core masses in Setal02's Fig. 6 are based on single-dish (beamsize
$11{''}$) observations, IRAS\,05506's mm-continuum core mass likely compares well with even
higher core-mass sources in Setal02's Fig. 6, which has 3 more sources with core masses less
than about 30\ms. We note that the HMPO list of Setal02 may well include
stars at the upper end of the intermediate-mass range (i.e., 8\,\ms); for example, we
have found at least one example of an HMPO from this list, IRAS\,20126+4104, for which
recent observations suggest a mass of about 7\ms~(Cesaroni et al. 2005).

The optical detection of IRAS\,05506 might seem surprising, considering the general lack of
optical detections of HMPOs. 
It is plausible that the outflows from IRAS\,05506 have cleared away enough
circumstellar cloud material from the near-vicinity of the central star, allowing a
detectable amount of starlight at optical wavelengths to leak out and be scattered into
our line-of-sight by dust in the walls of the cavity cleared out by the
outflow. Note that this scenario does not
require the outflow cavity to be directed exactly along the line-of-sight. In addition, 
we do not know the inclination and opening angles of the molecular outflow, and it is
quite possible that these are sufficiently favorable for the above scenario to be
feasible.

The jet-like feature emanating from the central source is directed along an axis roughly
orthogonal to the average position angle
for the most prominent knots. This suggests that either the collimated outflow
which created the knots has suffered a drastic change in direction, or {\it Sa-jet} is the
result of an
independent high-velocity outflow. If the latter hypothesis is correct, it could 
indicate the presence of an unresolved binary star associated with source {\it Sa}.
Taking the lowest value (3150\ls) of the total luminosity as derived from the SED
modelling (\S\,\ref{dust_phys}), and dividing it equally between the two stars in the
binary, we find that
these stars lie on the Bernasconi \&
Maeder (1996) pre-main-sequence tracks for 9\ms~stars. Hence, even if the central star is
a binary, at least one of the components is a massive star (i.e., if D is 2.2\,kpc or
larger). Note that the expansion time-scales derived for the fast molecular outflow and
the fast-moving optical knots are very similar, indicating that these were initiated
quasi-simultaneously in IRAS\,05506's recent history. So if the optical and molecular
outflows are produced from different stars in a binary system, their similar expansion
time-scales suggest a common triggering mechanism for launching them.

\section{Conclusions}
We have made a serendipitous discovery of an enigmatic outflow source, IRAS\,05506+2414,
as part of a multi-wavelength survey of pre-planetary nebulae. Subsequent to the HST
imaging, we carried out optical spectroscopy, millimeter-wave interferometry of the CO
J=1--0 line, as well as water and methanol maser line and ammonia inversion line
observations at $\sim(22-25)$\,GHz.

\noindent (1) The HST optical and near-infrared images show a bright compact central
source, and a fan-like spray of high-velocity (projected radial velocity 350\,\kms)
elongated knots which appear to emanate from it. These structures appear to be analogous
to the near-IR ``bullets" around the Orion KL nebula and to have similar masses
($\sim10^{-5}$\,\ms). The central source shows a jet-like extension along
PA$\sim$230\arcdeg, which is roughly orthogonal to the average direction of the bullet spray  
seen in the optical images.

\noindent (2) The high-resolution (3\farcs9$\times$2\farcs9) CO J=1--0 data show a compact
emission core of size ($\lesssim$ 5\arcsec); the line profile has a narrow core and weak,
high-velocity wings. The latter represent emission from a high-velocity bipolar outflow
directed along the optical jet feature. A 10\,mJy continuum source (at $\lambda=2.6$\,mm)
is also seen at the same location as the line peak. The continuum source structure
consists of an unresolved core and a faint extension whose orientation is consistent with
the optical jet and high-velocity molecular outflow direction.

\noindent (3) H$_2$O maser emission was detected both with the VLA and the GBT. With the
latter, NH$_3$\,(1,1) and (2,2) lines were detected, but no methanol emission was found
(rms antenna temperature was $\sim$25\,mK). The H$_2$O maser emission is spatially
unresolved. Our map of NH$_3$ emission shows an extended molecular cloud of size about
1$'$, with a kinetic temperature of 12.5\,K, and a mass of about 11\ms, assuming a
fractional NH$_3$/H$_2$ abundance ratio (by number), $f(NH_3)=4\times10^{-8}$, and a
distance of 2.8\,kpc. In comparison, the virial mass of the cloud is 48(73)\,\ms, for a
r$^{-2}$ (r$^{-1}$) density distribution. The factor 4 (or more) discrepancy between the
virial cloud mass and that derived from the ammonia data may be attributed to the large
uncertainty in our assumed value of $f(NH_3)$ and the distance, and/or to
non-gravitational linewidths, such as unresolved outflow motion.

\noindent (4) The 8\,GHz continuum observations show no continuum emission (3$\sigma$
upper limit of 150 $\mu$Jy) within a field of approximately 40\arcsec~diameter centered on
the position of IRAS\,05506.

\noindent (5) IRAS\,05506 is most likely not a PPN. The combined presence of H$_2$O and OH
maser, far-IR, NH$_3$, CO, CS and C$^{34}$S emission strongly argues for a dense, 
star-forming core being associated with IRAS\,05506. IRAS\,05506 is most likely either a
massive or intermediate-mass YSO, but not a low-mass YSO. If it is a massive star then the
lack of radio continuum and the late G -- early K spectral type we find for the central
star implies that in IRAS\,05506, we are witnessing the earliest stages of a massive
star's life, while its temperature is still too low to provide sufficient UV flux for
ionisation.

\noindent (6) IRAS\,05506 resembles objects in a recently identified class of high-mass
protostellar objects (HMPOs). Given the relatively low gas kinetic temperature and
narrow line-width derived from the ammonia data, IRAS05506 could be an extreme member of this
class, lying at the lower end of the mass range for these objects.

\noindent (7) A multi-component dust model provides a reasonable fit to the mid-to-far
infrared SED of IRAS\,05506; we find that the bulk of the mass is in a cool dust
component with a temperature of 27\,K and a total (gas+dust) mass of 26\,\ms (and about 1\% of
this in a warmer component at 79\,K), assuming a
gas-to-dust ratio of 100. The 2.6\,mm continuum flux measured with OVRO gives a total
mass of 2.2--6.4\,\ms in a $\sim$4$''$-size region around source {\it Sa}, assuming dust
temperatures in the range 79--27\,K.

\noindent (8) Both the molecular high-velocity outflow and that producing the optical
knots have extremely short and similar expansion time-scales of about 200\,yr, suggesting
that either these are manifestations of a common ejection event in IRAS05506's recent
past, or if they are due to independent ejections, there must have been a common
triggering mechanism for launching them. The molecular outflow cannot be driven by
radiation pressure since the expansion time-scale is orders of magnitude smaller than the
time required by radiation pressure to accelerate the observed bipolar outflow to its
current speed.

\noindent (9) We have estimated physical parameters of the high-velocity bipolar outflow
from the interferometric CO data -- our estimates are lower limits because the CO
emission is optically thick. We find that the outflow mass is, $M_{bip}>0.18$\ms, the
mass-outflow rate is, \mdot$_{bip}>9.7\times10^{-4}$\my, the scalar momentum is,
$P_{sc}>4.3$\,\ms\,\kms, the mechanical energy is, $E_{mech}>4.6\times10^{45}$ erg, and
the mechanical power is, $L_{mech}>7.9\times10^{35}$ erg s$^{-1}$.

\acknowledgments
We thank an anonymous referee for his/her comments which helped improve an earlier version of
this paper. RS and MM thank NASA for partially funding this work by a
NASA LTSA award (no. 399-20-40-06); RS also received partial support for
this work from HST/GO awards (nos.
GO-09463.01-A and GO-09801.01-A) from the Space Telescope Science Institute
(operated by the Association of Universities for Research in Astronomy,
under NASA contract NAS5-26555). CSC was partially funded for this work by
National Science Foundation grant 9981546 to Owens Valley Radio Observatory; 
the Spanish MCyT under project AYA2006-14876, the Spanish MEC under project PIE 200750I028, and
the Astrocam project (Ref: S-0505 ESP-0237).

\clearpage

\clearpage

\clearpage

\clearpage

\clearpage
                                                                                                     
\clearpage

\clearpage

\clearpage

\clearpage

\clearpage

\clearpage
                                                                                          
\end{document}